\documentclass[12pt]{article}
\usepackage{enumerate}
\usepackage{amsfonts}
\usepackage{amsmath}
\usepackage{amssymb}
\usepackage{amsthm}

\def\H{\mathcal{H}}
\def\K{\mathcal{K}}
\def\S{\mathfrak{S}}
\def\T{\mathfrak{T}}
\def\B{\mathfrak{B}}
\def\P{\mathcal{P}}

\newcommand{\id}{\mathrm{Id}}
\newcommand{\sr}{\mathrm{SR}}
\newcommand{\sn}{\mathrm{SN}}
\newcommand{\Tr}{\mathrm{Tr}}

\newcommand{\es}{\mathrm{ess}\,\mathrm{sup}}

\newcounter{defin}  \newcounter{lemma}  \newcounter{theorem}
\newcounter{property} \newcounter{corol}  \newcounter{remark} \newcounter{example}

\newenvironment{lemma}{\par\refstepcounter{lemma}%\noindent
     \textbf{Lemma \thelemma.} }{\rm\par}

\newenvironment{property}{\par\refstepcounter{property}%\noindent
     \textbf{Proposition \theproperty.}\ }{\rm\par}
\newenvironment{corollary}{\par\refstepcounter{corol}%\noindent
     \textbf{Corollary \thecorol.} }{\rm\par}

\begin{document}
\title{The Schmidt number and partially entanglement breaking channels in infinite dimensions\footnote{This work is partially supported by the program
"Mathematical control theory" of Russian Academy of Sciences, by the
federal target program "Scientific and pedagogical staff of
innovative Russia" (program 1.2.1, contract P 938), by the
analytical departmental target program "Development of scientific
potential of the higher school" (project 2.1.1/500) and by
RFBR grants 09-01-00424-a and 10-01-00139-a.}}
\author{M.E. Shirokov\\
Steklov Mathematical Institute, RAS, Moscow\\
msh@mi.ras.ru}
\date{}
\maketitle

\begin{abstract}
A definition of the Schmidt number of a state of an infinite
dimensional bipartite quantum system is given and properties of the
corresponding family of Schmidt classes are considered.  The
existence of states with a given Schmidt number such that any their
countable convex decomposition does not contain pure states with
finite Schmidt rank is established.

Partially entanglement breaking channels in infinite dimensions are
studied. Several properties of these channels well known in finite
dimensions are generalized to the infinite dimensional case. At the
same time, the existence of partially entanglement breaking channels
(in particular, entanglement breaking channels) such that all
operators in any their Kraus representations have infinite rank is
proved.
\end{abstract}
\maketitle

\pagebreak
\tableofcontents

\section{Introduction}

The Schmidt rank of a pure state and its extension to mixed states
called the Schmidt number are important quantitative characteristics
of entanglement in bipartite quantum systems.

The Schmidt rank of a pure state of bipartite system $AB$ determined
by a unit vector $|\psi\rangle$ is defined as the number of nonzero terms in
the Schmidt decomposition
$$
|\psi\rangle=\sum_i\lambda_i |\alpha_i\rangle\otimes|\beta_i\rangle
$$
of this vector, it coincides with the rank of the reduced states.

The Schmidt number of a mixed state $\omega$ of a finite dimensional
bipartite system $AB$ is defined in \cite{T&H} as the minimum over
all its decompositions into convex combination of pure states of the
maximal Schmidt rank of these pure states (see Section 3). In
\cite{T&H} it is shown that the Schmidt number does not increase
under LOCC-operations and that the set of states with the Schmidt
number not exceeding $k$ (the Schmidt class of order $k$) can be
characterized in terms of $k$-positive maps. In the subsequent
papers \cite{SBL,S&V-1,S&V-2} different properties of the Schmidt
number and of the family of the Schmidt classes are considered. In
particular, the notion of  $k$-Schmidt witnesses generalizing the
notion of entanglement witnesses is introduced and analyzed.

By using the Schmidt number the notion of partially entanglement
breaking quantum channels is introduced in \cite{p-e-b-ch}. It turns out
that this notion is closely related to the necessary condition of
nondecreasing of the Holevo quantity of an ensemble of quantum
states under action of a quantum channel \cite[Theorem 1]{Sh-14}.

This paper is devoted to infinite-dimensional generalizations of the
above concepts.  It is partially motivated by author's intension to
extend the above-mentioned result in \cite{Sh-14} to "continuous
variable" systems and channels.

In Section 3  a natural generalization of the Schmidt number to
states of infinite dimensional bipartite quantum systems is
considered. Since existence of non-countably decomposable separable
states (see \cite{H-Sh-W}) makes the finite dimensional formula for the
Schmidt number non-adequate, the "continuous" modification of this
formula (based on the notion of the essential supremum of a function
with respect to a given measure) is proposed. It is shown that this
formula gives a reasonable definition of the Schmidt number in the
sense that the corresponding Schmidt classes (the sets of states
with the Schmidt number  $\leq k$) coincide with the convex closures
of the sets of pure states with the Schmidt rank $\leq k$.

Properties of the Schmidt classes in infinite
dimensions are considered in Section 4. In particular, the
characterization the Schmidt class of order $k$ in terms of
$k$-positive maps (generalizing Theorem 1 in \cite{T&H}) is given.
It is shown that an arbitrary state in the Schmidt class of order
$k$ can be represented as a barycenter of a probability measure
supported by pure states with the Schmidt rank $\leq k$. At the same
time, the existence of states with a given Schmidt number such that
any their countable convex decomposition does not contain pure
states with a finite Schmidt rank is established.

A definition and some properties of partially entanglement breaking
channels in infinite dimensions are considered in Section 5. In
contrast to the finite dimensional case,  the class of
$k$\nobreakdash-\hspace{0pt}partially entanglement breaking channels
does not coincide with the class of channels having the Kraus
representation consisting of operators of rank $\leq k$ (the latter
is a proper subclass of the former). Moreover, the existence of
partially entanglement breaking channels (in particular, of
entanglement breaking channels) such that all operators in any their
Kraus representations have infinite rank is proved.

\section{Basic notations}

We will use the following notations:

\begin{description}
  \item[$\mathcal{H}, \mathcal{H}', \mathcal{K}$] -- separable Hilbert spaces;
  \item[$\mathfrak{B}(\mathcal{H})$] -- the Banach space of all bounded
operators in $\mathcal{H}$;

%\item[$\mathfrak{B}_{\mathrm{h}}(\mathcal{H})$] -- the real Banach space of all bounded Hermitian operators in $\mathcal{H}$;

  \item[$\mathfrak{T}( \mathcal{H})$] -- the Banach space of all
trace\nobreakdash-\hspace{0pt}class operators in $\mathcal{H}$;

  \item[$\mathfrak{T}_+( \mathcal{H})$]  -- the cone
of all positive
trace\nobreakdash-\hspace{0pt}class operators in $\mathcal{H}$;

 \item[$\mathfrak{S}(\mathcal{H})$]  -- the subset of $\mathfrak{T}_+( \mathcal{H})$ consisting of operators with a unit trace.
\end{description}

The unit operator in a Hilbert space $\mathcal{H}$ and the identity
transformation of the space $\mathfrak{T}(\mathcal{H})$ are denoted
$I_{\mathcal{H}}$ and $\mathrm{Id}_{\mathcal{H}}$ correspondingly.

Operators in $\mathfrak{S}(\mathcal{H})$ are denoted
$\rho,\sigma,\omega,...$ and called density operators or states
since each density operator uniquely defines a normal state on
$\mathfrak{B}(\mathcal{H})$.

We denote by $\mathrm{cl}(\mathcal{A})$, $\mathrm{co}(\mathcal{A})$,
$\overline{\mathrm{co}}(\mathcal{A})$ and
$\mathrm{extr}(\mathcal{A})$ the closure, the convex hull, the
convex closure and the set of all extreme points of a subset
$\mathcal{A}$ of a linear topological space correspondingly
\cite{J&T,ES,R}.

The set of all Borel probability measures on a closed subset
$\mathcal{A}\subseteq\mathfrak{S}(\mathcal{H})$ endowed with the
topology of weak convergence is denoted $\mathcal{P}(\mathcal{A})$ \cite{Bill,Par}.
This set can be considered as a complete separable metric space
\cite{Par}. The barycenter $\textbf{b}(\mu)$ of a measure
$\mu$ in $\mathcal{P}(\mathcal{A})$ is the state in
$\overline{\mathrm{co}}(\mathcal{A})$ defined by the Bochner
integral
\[
\textbf{b}(\mu)=\int_{\mathcal{A}}\rho\mu(d\rho).
\]

For an arbitrary subset
$\,\mathcal{B}\subseteq\overline{\mathrm{co}}(\mathcal{A})\,$ let
$\mathcal{P}_{\mathcal{B}}(\mathcal{A})$ be the subset of
$\mathcal{P}(\mathcal{A})$ consisting of all measures with the
barycenter in $\mathcal{B}$.

A finite or infinite collection of states
$\{\rho_{i}\}\subset\mathcal{A}\subseteq\mathfrak{S}(\mathcal{H})$
with the corresponding probability distribution $\{\pi_{i}\}$ is
conventionally called \textit{ensemble} and denoted
$\{\pi_{i},\rho_{i}\}$. An ensemble can be considered as an atomic
(discrete) measure in $\mathcal{P}(\mathcal{A})$. The barycenter of
this measure is the average state $\sum_{i}\pi_{i}\rho_{i}$ of the
corresponding ensemble.

A  positive trace\nobreakdash-\hspace{0pt}preserving linear map
$\Phi:\mathfrak{T}(\mathcal{H})\rightarrow\mathfrak{T}(\mathcal{H}')$
such that the dual map
$\Phi^{*}:\mathfrak{B}(\mathcal{H}')\rightarrow\mathfrak{B}(\mathcal{H})$
is completely positive is called  \textit{quantum channel}
\cite{H-SSQT,N&Ch}. The set of all quantum channels from
$\mathfrak{T}(\mathcal{H})$ to $\mathfrak{T}(\mathcal{H}')$ is denoted $\mathfrak{F}(\mathcal{H},\mathcal{H}')$.

\section{The Schmidt number}

Let $\H$ and $\K$ be separable Hilbert spaces. The Schmidt rank
$\sr(\omega)$ of a pure state $\omega$ in $\S(\H\otimes\K)$ can be
defined as the rank of the isomorphic states $\Tr_{\K}\omega$ and
$\Tr_{\H}\omega$.

If the spaces $\H$ and $\K$ are finite dimensional then the Schmidt
number of an arbitrary state $\omega$ in $\S(\H\otimes\K)$ is
defined in \cite{T&H} as follows
\begin{equation}\label{sn-def-1}
    \sn(\omega)=\inf_{\sum_i \pi_i\omega_i=\omega}\sup_i \sr(\omega_i)
\end{equation}
(the infimum is over all ensembles $\{\pi_i,\omega_i\}$ of pure
states with the average state $\omega$). By using the Caratheodory
theorem it is easy to show that for each natural $k$ the set
$\S_k(\H\otimes\K)=\{\omega\in\S(\H\otimes\K)\,|\,\sn(\omega)\leq k\}$ is compact and coincides
with the convex hull of all pure states having the Schmidt rank
$\leq k$. This implies that the function $\omega\mapsto\sn(\omega)$ is
lower semicontinuous on the set $\S(\H\otimes\K)$. Thus we have the
increasing finite sequence\footnote{Here and in what follows we will often write $\S_k$ instead of $\S_k(\H\otimes\K)$ for brevity.}
$$
\S_1\subset\S_2\subset\S_3\subset...\subset\S_{n-1}
\subset\S_n=\S(\H\otimes\K)
$$
of compact sets, where $\S_1$ is the
set of separable (non-entangled) states and
$n=\min\{\dim\H,\dim\K\}$.

If the spaces $\H$ and $\K$ are infinite dimensional then the right
side of (\ref{sn-def-1}) is well defined but can not be used as an
adequate definition of the Schmidt number. This follows from existence
of separable states in $\S(\H\otimes\K)$
(called non-countably decomposable), which can not be decomposed into a countable convex combination of product pure states
\cite{H-Sh-W}. The nonexistence of a decomposition into product pure states shows that the right
side of (\ref{sn-def-1}) is $>1$ for any such
separable state $\omega$ contradicting to the natural requirement
for the Schmidt number.\footnote{This problem is similar to the
problem arising in infinite dimensional generalization of the convex
roof construction of entanglement monotones: the existence of
non-countably decomposable separable states makes the discrete
version of this construction not adequate (see Remark 9 in
\cite{Sh-9}).}

We will show below that a reasonable generalization of definition
(\ref{sn-def-1}) to the infinite dimensional case is given by the
following formula
\begin{equation}\label{sn-def-2}
    \sn(\omega)=\inf_{\mu\in\P_{\{\omega\}}(\mathrm{extr}\S(\H\otimes\K))}\es_{\mu} \sr(\cdot)
\end{equation}
where $"\es_{\mu}"$ means the essential supremum with respect to the
measure $\mu$ \cite[Section 13.1]{ES}. Note that $\es_{\mu} \sr(\cdot)=\|\sr\|_{\infty}$ -- the norm of the function $\sr$ in the space $L^{\infty}(X, \mu)$, where $X=\mathrm{extr}\S(\H\otimes\K)$. \medskip

\begin{property}\label{sn-def-p}
A) \emph{The function $\,\sn(\omega)$ defined by (\ref{sn-def-2}) is
lower semicontinuous on $\,\S(\H\otimes\K)$. For each state
$\,\omega\in\S(\H\otimes\K)$ the infimum in (\ref{sn-def-2}) is
achieved at some measure in
$\,\mathrm{extr}\P_{\{\omega\}}(\mathrm{extr}\S(\H\otimes\K))$.}\medskip

B)\emph{ For each natural $k$ the set
$\,\S_k(\H\otimes\K)=\{\omega\in\S(\H\otimes\K)\,|\,\sn(\omega)\leq
k\}$, where $\sn(\omega)$ is defined by (\ref{sn-def-2}), is closed
and convex. It coincides with the convex closure of all pure states
in $\,\S(\H\otimes\K)$ having the Schmidt rank $\leq k$.}\medskip

C) \emph{If $\,\omega$ is a finite rank state in $\,\S(\H\otimes\K)$
then the value $\,\sn(\omega)$ defined by (\ref{sn-def-2}) coincides
with the value $\,\sn(\omega)$ defined by (\ref{sn-def-1}).}\medskip
\end{property}

\textbf{Proof.} Since the nonnegative function
$\omega\mapsto\sr(\omega)$ is lower semicontinuous on
$\mathrm{extr}\S(\H\otimes\K)$, the first assertion of the
proposition follows from Proposition  \ref{appedix} in the Appendix.
\medskip

The second assertion follows from the first one and Lemma 1 in
\cite{H-Sh-W}. \medskip

To prove the third assertion assume that the right side of
(\ref{sn-def-2}) is equal to $k<+\infty$.  Equality of the right
side of (\ref{sn-def-1}) to $k$ follows from coincidence of the
convex hull and the convex closure of the closed subset
$$
\S^{\omega}_{k}=\{\varpi\in\mathrm{extr}\S(\mathrm{supp}\omega)\,|\,\sr(\varpi)\leq
k\}
$$
of the finite-dimensional space $\T(\mathrm{supp}\omega)$ (see
\cite[Corollary 5.33]{ES}). $\square$ \medskip

The following proposition generalizes  Proposition 1 in
\cite{T&H} to the infinite dimensional case.
\smallskip

\begin{property}\label{locc-monotonicity}
\emph{The Schmidt number of a state of an infinite-dimensional
bipartite system (defined by (\ref{sn-def-2})) does not increase
under LOCC-operations.}
\end{property}\smallskip

This proposition can be reduced to Proposition 1 in
\cite{T&H} by using the following approximation result.
\smallskip

\begin{lemma}\label{f-d-approx} \emph{Let $\{P_n\}$ and $\{Q_n\}$ be
increasing sequences of finite rank projectors strongly converging
respectively to $I_{\H}$ and to $I_{\K}$. For an arbitrary state
$\omega\in\S(\H\otimes\K)$ let $\,\omega_n=(\Tr P_n\otimes
Q_n\cdot \omega)^{-1}P_n\otimes Q_n\cdot \omega \cdot P_n\otimes
Q_n$. Then}
$$
\lim_{n\rightarrow+\infty}\sn(\omega_n)=\sn(\omega).
$$
\emph{If $\,\sn(\omega)<+\infty$ then there exists $n_0$ such that
$\,\sn(\omega_n)=\sn(\omega)$ for all $n\geq n_0$.}
\end{lemma}
\smallskip
\textbf{Proof.} By lower semicontinuity of the Schmidt number (Proposition \ref{sn-def-p}A) it suffices to show that
\begin{equation}\label{l-m}
    \sn(\omega_n)\leq \sn(\omega),\quad \forall n.
\end{equation}

Since the state $\omega$ belongs to the convex closure of the set
$\S^p_{\sn(\omega)}$ of pure states with the Schmidt rank $\leq
\sn(\omega)$  (Proposition \ref{sn-def-p}B), there exists a sequence $\{\omega_m\}$ from the convex
hull of the set $\S^p_{\sn(\omega)}$ converging to the state
$\omega$ such that
$\lim_{n\rightarrow+\infty}\sn(\omega_m)=\sn(\omega)$. For each $m$
inequality (\ref{l-m}) with $\omega=\omega_m$ is directly
verified. By lower semicontinuity of the Schmidt number passing to
the limit as $m\rightarrow+\infty$ implies (\ref{l-m}). $\square$

\section{Some properties of the Schmidt classes  $\S_k$}

If $\,\dim\H=\dim\K=+\infty$ we have the increasing infinite
sequence
$$
\S_1\subset\S_2\subset\S_3\subset...\subset\S_{n-1}
\subset\S_n\subset...
$$ of closed convex subsets of $\S(\H\otimes\K)$, where $\S_1$ is the set
of separable (nonentangled) states.
\medskip

Let $\S^p_{k}$ be the closed subset of $\S_{k}$ consisting of pure
states.
\medskip
\begin{property}\label{int-rep} A) \emph{An arbitrary state in $\S_k$ can be represented as a
barycenter of some measure in
$\,\mathrm{extr}\P(\S^p_k)$.}\smallskip

B) \emph{There exist states $\omega$ in $\;\S_k\setminus\S_{k-1}\,$ such that the operator $\,\omega-\lambda\sigma$
is not positive for any $\lambda>0$ and any pure state $\sigma$ with
a finite Schmidt rank.\footnote{It is assumed that $\S_0=\emptyset$, so that $\S_1\setminus\S_0=\S_1$ is the set of separable states.} For any such state $\omega$ we have}
$$
\omega=\sum_i\pi_i
\omega_i,\quad\{\omega_i\}\subset\mathrm{extr}\S(\H\otimes\K),\qquad\Rightarrow\qquad\sr(\omega_i)=+\infty\quad\forall
i.
$$

C) \emph{An arbitrary pure state  in $\;\S_k\setminus\S_{k-1}\,$ can be approximated by states in $\;\S_k\setminus\S_{k-1}\,$ having the property stated in B).}

\end{property}
\medskip
\textbf{Proof.} The first assertion directly follows from
Proposition \ref{sn-def-p},  the second one -- from  the example in the Appendix 6.2 (after Proposition \ref{state}).

The third assertion can be derived from the construction in the example in the Appendix 6.2 by noting that functions with non-vanishing Fourier coefficients form a dense subset in $L^2([0,2\pi))$ and that an arbitrary set $\{|\psi_i\rangle\}_{i=1}^k$ of orthogonal unit vectors in a separable Hilbert space $\H$ can be represented as an image of the set $\{|\varphi_i\rangle\otimes|i\rangle\}_{i=1}^k\subset L^2([0,2\pi))\otimes\K$ under some unitary map from $L^2([0,2\pi))\otimes\K$ onto $\H$, where $\{|i\rangle\}_{i=1}^k$ is an orthonormal basis in the space $\K$.
$\square$\medskip

Consider the characterization of the set $\S_k$ in term of
$k$-positive maps (generalizing Theorem 1 in \cite{T&H} to the
infinite-dimensional case).
\medskip
\begin{property}\label{k-p-m} \emph{A state $\omega\in\S(\H\otimes\K)$ belongs to the set $\,\S_k$ if and
only if the operator $\Lambda_k\otimes\id_{\K}(\omega)$ is positive
for any  $k$-positive linear transformation $\,\Lambda_k$ of the space $\,\T(\H)$.}
\end{property}
\medskip
\textbf{Proof.} Let $\,\omega_0\in\S_k$. By Proposition
\ref{int-rep} there exists a measure $\,\mu_0$ in $\P(\S^p_k)$ such
that $\,\omega_0=\int \omega \mu_0(d\omega)$. Since
$\,\Lambda_k\otimes\id_{\K}(\omega)\geq0\,$ for any $\omega\in
\S^p_k$ by definition of $k$-positivity,
$\,\Lambda_k\otimes\id_{\K}(\omega_0)=\int
\Lambda_k\otimes\id_{\K}(\omega) \mu_0(d\omega)\geq0$.

The converse assertion can be derived from the corresponding
finite dimensional result (\cite[Theorem 1]{T&H})  by using the
approximation based on Lemma \ref{f-d-approx}.

Let $\omega_0\in\S(\H\otimes\K)\setminus\S_k$, i.e.
$\sn(\omega_0)>k$. By Lemma \ref{f-d-approx} there exist projectors $P\in\B(\H)$ and $Q\in\B(\K)$ of the same finite rank such that the
state $\omega_*=(\Tr P\otimes Q\cdot \omega_0)^{-1}P\otimes Q\cdot
\omega_0 \cdot P\otimes Q$ does not belong to the set $\S_k$. Let
$\H_*=P(\H)$ and $\K_*=Q(\K)$. By Theorem 1 in \cite{T&H} there
exists a $k$\nobreakdash-\hspace{0pt}positive map $\Lambda_k:\T(\H_*)\rightarrow\T(\H_*)$ such that the operator $\Lambda_k\otimes\id_{\K_*}(\omega_*)$
is not positive. Consider the $k$-positive map $\Lambda_k\circ\Pi$,
where $\Pi(\cdot)=P(\cdot)P$. Then the operator
$(\Lambda_k\circ\Pi)\otimes\id_{\K}(\omega_0)$ is not positive,
since  otherwise the operator
$$
I_{\H}\otimes Q
\cdot(\Lambda_k\circ\Pi)\otimes\id_{\K}(\omega_0)\cdot I_{\H}\otimes
Q=(\Tr P\otimes Q\cdot \omega_0)
\,\Lambda_k\otimes\id_{\K_*}(\omega_*)
$$
is positive in contradiction to the choice of $\Lambda_k$. $\square$
\medskip

By using the compactness criterion for subsets of
$\T_{+}(\H\otimes\K)$ (see the Proposition in the Appendix in \cite{Sh-H}) one can generalize Proposition 1 in \cite{SBL} to infinite dimensions. \medskip

\begin{property}\label{dec-p} \emph{An arbitrary state $\omega_k\in\S_k$ can be represented as follows
\begin{equation}\label{dec}
\omega_k=(1-p)\omega_{k-1}+p\delta,\quad p\in[0,1],
\end{equation}
where $\omega_{k-1}\in\S_{k-1}$ and $\delta$ is a state having the
Schmidt number $\geq k$ such that the operator
$\,\delta-\lambda\sigma$ is not positive for any
$\,\lambda>0$ and any $\,\sigma\in\S_{k-1}$.}

\emph{Among all such decompositions there is a decomposition with
minimal $\,p$.}
\end{property} \smallskip

A state having the property of the state $\delta$  is called $k$-\emph{edge} state in  \cite{SBL}.
In contrast to the finite dimensional case, to prove that $\delta$ is a $k$-edge state it is not sufficient to show that $\,\delta-\lambda\sigma$ is not positive for any
$\,\lambda>0$ and any $\,\sigma\in\S^p_{k-1}$. This follows form Proposition \ref{int-rep}B.  \smallskip

\textbf{Proof.} Let
$\mathcal{M}=\{0\}\cup\{A\in\T_{+}(\H\otimes\K)\,|\,A\leq\omega_k, (\Tr
A)^{-1}A\in\S_{k-1}\}$ be a closed subset of the cone
$\T_{+}(\H\otimes\K)$.

Assume $\mathcal{M}\neq\{0\}$. By the above-mentioned compactness criterion for subsets of
$\T_{+}(\H\otimes\K)$ the set $\mathcal{M}$ is compact. Hence
there exists $A_0\in \mathcal{M}$ such that $\Tr A_{0}=\sup_{A\in\mathcal{M}}\Tr A$.
Denoting  $p=1-\Tr A_{0}$,  $\omega_{k-1}=(\Tr A_{0})^{-1}A_{0}$ and
$\delta=p^{-1}(\omega_k-A_0)$ we obtain (\ref{dec}) with minimal $p$.

If $\mathcal{M}=\{0\}$ then the only way to obtain (\ref{dec}) is to take
$p=1$ and $\delta=\omega_k$. $\square$\medskip

Since the family $\{\S_k\}$ consists of  closed convex subsets of
the Banach space $\T_h(\H\otimes\K)$ of all trace class Hermitian
operators in $\H\otimes\K$, for each
$k$ and each $\omega_0\in \S_k\setminus\S_{k-1}$ the Hahn-Banach theorem implies existence of a
Hermitian operator $W$ such that $\Tr W\omega\geq 0$ for all
$\omega\in\S_{k-1}$ and $\Tr W\omega_0< 0$. This operator $W$ is
called  \emph{$k$-Schmidt witness ($k$-SW) detecting} $\omega_0$
\cite{SBL}.

An explicit form of $k$-SW detecting a state $\delta$ such that the
operator $\,\delta-\lambda\sigma$ is not positive for any
$\,\lambda>0$ and $\,\sigma\in\S_{k-1}$ (in particular, the
$k$-edge state $\delta$ in decomposition (\ref{dec})) is given in
\cite[Lemma 1]{SBL} (in finite dimensions). This is the operator
$$
W=P-\frac{\varepsilon}{c}C,
$$
where $P$ is an arbitrary positive operator whose range coincides
with the kernel of the state $\delta$, $C$ is an arbitrary positive
operator such that $\Tr C\delta>0$,
$\varepsilon=\inf_{|\varphi\rangle\langle\varphi|\in\S^p_{k-1}}
\langle\varphi|P|\varphi\rangle$ and $c=\|C\|$ (compactness
arguments imply $\varepsilon>0$).\smallskip

To apply the above construction in infinite dimensions we have to
impose additional conditions ensuring positivity of $\varepsilon$.
\smallskip
\begin{property}\label{k-sw} \emph{The above construction of  $k$-SW detecting the $k$-edge state $\delta$  is valid in infinite dimensions
if the state $\delta$ has finite rank and the spectrum of the
operator $P$ does not contain zero.}
\end{property} \medskip

\textbf{Proof.} By Proposition \ref{int-rep}A to prove this assertion it suffices to show that
$\varepsilon=\inf_{|\varphi\rangle\langle\varphi|\in\S^p_{k-1}}
\langle\varphi|P|\varphi\rangle>0$ if $P$ is the projector on the
kernel of $\delta$.

Assume that $\varepsilon=0$. Then there exists a sequence
$\{|\varphi_n\rangle\}$ of  vectors such that
$|\varphi_n\rangle\langle\varphi_n|\in \S^p_{k-1}$ for each $n$ and
$\lim_{n\rightarrow+\infty}\langle\varphi_n|I_{\H\otimes\K}-P|\varphi_n\rangle=1$.
The weak compactness of the unit ball of $\H\otimes\K$ implies existence
of a subsequence $\{|\varphi_{n_k}\rangle\}$ weakly converging to a vector
$|\varphi_*\rangle$. Since $I_{\H\otimes\K}-P$ is a finite rank projector,
we have
$$
\langle\varphi_*|I_{\H\otimes\K}-P|\varphi_*\rangle=\lim_{n\rightarrow+\infty}\langle\varphi_n|I_{\H\otimes\K}-P|\varphi_n\rangle=1.
$$
Thus $|\varphi_*\rangle$ is a unit vector belonging to the range of $\delta$
and hence the subsequence $\{|\varphi_{n_k}\rangle\}$ converges to the
vector $|\varphi_*\rangle$ in the norm topology. This implies
$|\varphi_*\rangle\langle\varphi_*|\in \mathfrak{S}^p_{k-1}$ contradicting to
the basic property of the state $\delta$. $\square$

\section{Partially entanglement breaking channels}

The notion of a $k$-partially entanglement-breaking ($k$-PEB)
channel in finite dimensions is introduced in \cite{p-e-b-ch} as a
natural generalization of the notion of an entanglement-breaking
channel (which is $1$-PEB). According to \cite{p-e-b-ch},
\emph{a channel $\,\Phi:\T(\H)\rightarrow\T(\H')$ is called $k$-partially
entanglement-breaking if for an arbitrary Hilbert space $\K$ the
Schmidt number of the state $\,\Phi\otimes\id_{\K}(\omega)\in\S(\H'\otimes\K)$ does not
exceed $k$ for any state $\omega\in\S(\H\otimes\K)$}.

By using the definition of the Schmidt number introduced in Section
3 the above definition of $k$-partially entanglement breaking
channels is directly generalized to the infinite-dimensional case.

Denote by $\mathfrak{P}_k(\H,\H')$ the class of $k$-partially
entanglement breaking channels from $\T(\H)$ to $\T(\H')$. Since the set $\S_k(\H'\otimes\K)$ is closed and convex,  $\mathfrak{P}_k(\H,\H')$ is a closed convex subset of the set $\mathfrak{F}(\H,\H')$ of all channels from $\T(\H)$ to  $\T(\H')$ endowed with the strong convergence topology \cite{Sh-H}.

%In what follows we will often write for brevity $\mathfrak{P}_k$ instead of $\mathfrak{P}_k(\H,\H')$.

Proposition \ref{k-p-m}  implies the following characterization of $k$-PEB channels (generalizing the corresponding finite-dimensional results \cite{p-e-b-ch,e-b-ch}).
\medskip
\begin{property}\label{peb-c} \emph{A channel $\,\Phi$ is $k$-partially entanglement breaking if and only if
the map $\Lambda_{k}\circ\Phi$ is completely positive for any
$k$-positive map $\Lambda_{k}$.}
\end{property}
\medskip

By
definition $\Phi\in\mathfrak{P}_k(\H,\H')$ means that
$\Phi\otimes\id_{\K}(\omega)\in\S_k(\H'\otimes\K)$ for any $\omega\in
\S(\H\otimes\K)$. By the following proposition it suffices to verify the
above inclusion only for one pure state.\medskip

\begin{property}\label{peb-sc}
\emph{Let $\,\Phi:\T(\H)\rightarrow\T(\H')$ be a quantum channel. If there exists a pure state $|\psi\rangle\langle\psi|$ in
$\S(\H\otimes\K)$ having full rank partial states
$\Tr_{\K}|\psi\rangle\langle\psi|\cong\Tr_{\H}|\psi\rangle\langle\psi|$
such that
$\,\Phi\otimes\id_{\K}(|\psi\rangle\langle\psi|)\in\S_k(\H'\otimes\K)$
then the channel $\,\Phi$ is $k$-partially
entanglement-breaking.}\medskip
\end{property}

\textbf{Proof.} Let $|\psi\rangle=\sum_{i=1}^{+\infty}\mu_i|i\rangle\otimes
|i\rangle$, where $\{|i\rangle\}$ is an orthonormal basis in
$\H\cong\K$ and $\mu_i>0$ for all $i$. Let $P_n=\sum_{i=1}^n
|i\rangle\langle i|\,$ be a projector in $\B(\K)$.

By Proposition \ref{locc-monotonicity} we have
$$
\Phi\otimes\id_{\K}(|\psi_n\rangle\langle\psi_n|)=c_n I_{\H}\otimes
P_n\cdot\Phi\otimes\id_{\K}(|\psi\rangle\langle\psi|)\cdot
I_{\H}\otimes P_n\in\S_k(\H'\otimes\K)
$$
where $|\psi_n\rangle=c_n\sum_{i=1}^{n}\mu_i|i\rangle\otimes
|i\rangle$ and $c_n=[\sum_{i=1}^{n}\mu_i^2]^{-1/2}$. \smallskip

Let $\H_n=\mathrm{lin}(\{|i\rangle\}_{i=1}^n)$ and
$\K_n=\mathrm{lin}(\{|i\rangle\}_{i=1}^n)$ be $n$-dimensional
subspaces of $\H$ and $\K$.  An arbitrary vector $|\varphi\rangle$
in $\H_n\otimes\K_n$ can be represented as follows
$|\varphi\rangle=\sum_{i,j=1}^n \gamma_{ij}|i\rangle\otimes
|j\rangle=\sum_{i=1}^n \mu_i |i\rangle\otimes A|i\rangle$, where
$A=\sum_{i,j=1}^n (\mu_i)^{-1}\gamma_{ij}|j\rangle\langle i|$ is an
operator in $\B(\K_n)$. Thus
$|\varphi\rangle\langle\varphi|=I_{\H_n}\otimes A
\cdot|\psi_n\rangle\langle\psi_n|\cdot I_{\H_n}\otimes A^*$ and
hence
$$
\Phi\otimes\id_{\K}(|\varphi\rangle\langle\varphi|)=I_{\H}\otimes
A\cdot \Phi\otimes\id_{\K}(|\psi_n\rangle\langle\psi_n|)\cdot
I_{\H}\otimes A^*\in\S_k(\H'\otimes\K).
$$
This means that the
restriction of the channel $\,\Phi$ to the set $\S(\H_n)$ is
$k$\nobreakdash-\hspace{0pt}partially entanglement breaking. By the below Lemma
\ref{peb-sc-2} the channel $\,\Phi$  is $k$\nobreakdash-\hspace{0pt}partially
entanglement breaking. $\square$ \medskip

The following lemma reduces a proof of the $k$\nobreakdash-\hspace{0pt}partially
entanglement breaking property of  a channel
$\,\Phi:\T(\H)\rightarrow\T(\H')\,$ to analysis of its finite\nobreakdash-\hspace{0pt}dimensional restrictions. \smallskip

\begin{lemma}\label{peb-sc-2}
\emph{Let $\{\H_n\}$ be an increasing sequence of subspaces
of $\H$ such that  $\,\mathrm{cl}(\bigcup_n\H_n)=\H$. If the
restriction of the channel $\,\Phi$ to the set $\,\S(\H_n)$ is
$k$\nobreakdash-\hspace{0pt}partially entanglement breaking for each $\,n\,$ then the channel $\,\Phi$ is
$k$\nobreakdash-\hspace{0pt}partially entanglement breaking.}\medskip
\end{lemma}

\textbf{Proof.} Since an arbitrary state $\omega\in\S(\H\otimes\K)$ can be approximated by a sequence $\{\omega_n\}$ such that $\mathrm{supp}\Tr_{\K}\omega_n\subset\H_n$ (see Lemma
\ref{f-d-approx}), this assertion follows from closedness of the set $\S_k(\H'\otimes\K)$. $\square$ \medskip

Let $|\psi\rangle\langle\psi|$ be a pure state in
$\S(\H\otimes\K)$ having full rank partial states
$\Tr_{\K}|\psi\rangle\langle\psi|\cong\Tr_{\H}|\psi\rangle\langle\psi|=\sigma$. Consider the Choi-Jamiolkowski  one-to-one correspondence
$$
\mathfrak{F}(\H,\H')\ni\Phi\leftrightarrow\Phi\otimes\id_{\K}(|\psi\rangle\langle\psi|)
\in\mathfrak{C}_{\sigma}\doteq\{\omega\in\S(\H'\otimes\K)\,|\,\Tr_{\H'}\omega=\sigma\},
$$
which is a topological isomorphism provided the set $\mathfrak{F}(\H,\H')$ of all channels is endowed with the strong convergence topology \cite[Proposition 3]{Sh-H}. Proposition \ref{peb-sc} implies the following observation.
\smallskip
\begin{corollary}\label{C-J}
\emph{The restriction of the Choi-Jamiolkowski isomorphism to the class $\,\mathfrak{P}_k(\H,\H')$ is an isomorphism between this class and the closed subset $\,\S_k(\H'\otimes\K)\cap\mathfrak{C}_{\sigma}$ of the set $\,\S(\H'\otimes\K)$.}

\emph{The set $\,\mathfrak{P}_k(\H,\H')\setminus\mathfrak{P}_{k-1}(\H,\H')$ corresponds to the set
$$
(\S_k(\H'\otimes\K)\setminus\S_{k-1}(\H'\otimes\K))\cap\mathfrak{C}_{\sigma},\quad k=2,3,...
$$
under this isomorphism.}
\end{corollary}\medskip

In \cite{p-e-b-ch} it is proved that a
channel $\Phi$ is $k$-PEB if and only if it has the Kraus
representation
\begin{equation}\label{k-r}
 \Phi(\cdot)=\sum_{i}V_i(\cdot)V_i^*
\end{equation}
such that $\mathrm{rank}V_i\leq k$ for all $\,i$ (this is a natural generalization of the well known
characterization of entanglement breaking finite-dimensional channels proved in
\cite{e-b-ch}). In infinite dimensions the class of  $k$-PEB channels is wider than the class of channels having the last property.\smallskip

\begin{property}\label{Kraus-rep} A) \emph{A channel $\,\Phi$ belongs to the class $\,\mathfrak{P}_k(\H,\H')$  if it has the Kraus representation
(\ref{k-r}) such that $\,\mathrm{rank}V_i\leq k\,$  for all
$\,i$.}\smallskip

B) \emph{There exist channels $\,\Phi$ in
$\;\mathfrak{P}_k(\H,\H')\setminus\mathfrak{P}_{k-1}(\H,\H')\,$ with the following property}\footnote{It is assumed that $\mathfrak{P}_0(\H,\H')=\emptyset$, so that $\mathfrak{P}_1(\H,\H')\setminus\mathfrak{P}_{0}(\H,\H')=\mathfrak{P}_1(\H,\H')$ is the class of entanglement breaking channels.}
$$
\Phi(\cdot)=\sum_iV_i(\cdot)V_i^*\qquad\Rightarrow\qquad\mathrm{rank}V_i=+\infty\quad\forall\, i.
$$
\end{property}

\textbf{Proof.} The first assertion is obvious, since for an arbitrary pure state $\omega\in\S(\H\otimes\K)$ the expression
$\Phi\otimes\id_{\K}(\omega)=\sum_iV_i\otimes I_{\K}\cdot\omega\cdot V_i^*\otimes I_{\K}$ gives a decomposition of the state $\Phi\otimes\id_{\K}(\omega)$ into a convex combination of pure states with the Schmidt rank $\leq k$.

To prove the second one consider a state $\,\omega$ in $\S_k(\H'\otimes\K)\setminus\S_{k-1}(\H'\otimes\K)$ having the property stated in Proposition \ref{int-rep}B. We may assume that $\Tr_{\H'}\omega$ is a full rank state in $\S(\K)$. Let $|\psi\rangle\langle\psi|$ be a purification of this state in $\S(\H\otimes\K)$. By Corollary \ref{C-J} the channel $\Phi_{\omega}$ corresponding to the state $\,\omega$ via the Choi-Jamiolkowski isomorphism induced by the state $|\psi\rangle\langle\psi|$ belongs to the set
$\mathfrak{P}_k(\H,\H')\setminus\mathfrak{P}_{k-1}(\H,\H')$. If we assume that $\Phi_{\omega}(\cdot)=\sum_iV_i(\cdot)V_i^*$ with $\mathrm{rank}V_{i_0}<+\infty$ for some $i_0$ then we will obtain a contradiction to the  property of the state $\,\omega$, since $V_{i_0}\otimes\id_{\K}|\psi\rangle\neq0$ (otherwise $V_{i_0}(\Tr_{\K}|\psi\rangle\langle\psi|)(V_{i_0})^*=0$ contradicting to a full rank of the state $\Tr_{\K}|\psi\rangle\langle\psi|$). $\square$
\medskip

\begin{corollary}\label{Kraus-rep} \emph{There exist entanglement breaking channels such that all operators in any their Kraus representations have infinite rank.}\smallskip
\end{corollary}

\section{Appendix}

\subsection{One property of the set $\S(\H)$.}

We consider here a corollary of the compactness
criterion for subsets of probability measures on the set $\S(\H)$
(described in detail in \cite[Section 1]{Sh-9}), which states that  \emph{a
subset $\P$ of $\P(\mathfrak{S}(\mathcal{H}))$ is compact (in the weak convergence topology) if and
only if  $\,\{\textbf{b}(\mu)\,|\,\mu\in\P\}$ is a compact subset of
$\,\mathfrak{S}(\mathcal{H})$ }.\medskip

\begin{property}\label{appedix}
\emph{Let $f$ be a nonnegative lower-semicontinuous function on a
closed subset $\mathcal{A}$ of $\,\S(\H)$. The function
\begin{equation}\label{ess-sup}
    F(\rho)=\inf_{\mu\in\P_{\{\rho\}}(\mathcal{A})}\es_{\mu} f(\cdot)
\end{equation}
is lower semicontinuous on the set
$\,\overline{\mathrm{co}}(\mathcal{A})$.\footnote{$"\es_{\mu}"$ means the essential supremum with respect to the
measure $\mu$ \cite[Sec.13.1]{ES}.} For each state
$\,\rho\in\overline{\mathrm{co}}(\mathcal{A})$ the infimum in
(\ref{ess-sup}) is achieved at some measure in
$\,\mathrm{extr}\P_{\{\rho\}}(\mathcal{A})$. }

\emph{For each  $c\geq 0$ the set $\,\{\rho\in
\overline{\mathrm{co}}(\mathcal{A})\,|\,F(\rho)\leq c\}$ coincides
with the convex closure of the set $\,\{\rho\in
\mathcal{A}\,|\,f(\rho)\leq c\}$. }
\end{property}\smallskip

\textbf{Proof.} Note first that the function
$F(\rho)$ is well defined on the set
$\,\overline{\mathrm{co}}(\mathcal{A})$ by Lemma 1 in \cite{H-Sh-W}.

Show that the functional
\begin{equation}\label{ess-sup-fun}
 \P(\mathcal{A})\ni\mu\mapsto \hat{f}(\mu)=\es_{\mu} f(\cdot)
\end{equation}
is concave and lower semicontinuous. Since for a given measure
$\mu\in\P(\mathcal{A})$ the $\mu$\nobreakdash-\hspace{0pt}essential supremum of the function
$f$ (coinciding with the norm $\|f\|_{\infty}$  in the space
$L^{\infty}(\mathcal{A},\mu)$) is the least upper bound of the
increasing family of the norms $\|f\|_{p}$ in the space
$L^{p}(\mathcal{A},\mu)$, $p\in[1,+\infty)$, concavity and lower
semicontinuity of functional (\ref{ess-sup-fun}) follow from concavity
and lower semicontinuity of the functional
\begin{equation*}
 \P(\mathcal{A})\ni\mu\mapsto \|f\|_{p}=\sqrt[p]{\int_{\mathcal{A}}
 [f(\rho)]^p\mu(d\rho)}
\end{equation*}
(lower semicontinuity of this functional follows from the basic properties of the weak convergence topology, see \cite[Chapter I, Sec.2]{Bill}).

By concavity and lower semicontinuity of functional (\ref{ess-sup-fun}) and
compactness of the set $\mathcal{P}_{\{\rho\}}(\mathcal{A})$
(provided by the above-stated compactness criterion) the infimum  in
the definition of the value $F(\rho)$ for each $\rho$ in
$\overline{\mathrm{co}}(\mathcal{A})$ is achieved at a particular
measure in $\mathrm{extr}\mathcal{P}_{\{\rho\}}(\mathcal{A})$.

Suppose function (\ref{ess-sup}) is not lower semicontinuous. Then there
exists a sequence
$\{\rho_{n}\}\subset\overline{\mathrm{co}}(\mathcal{A})$ converging
to a state $\rho_{0}\in\overline{\mathrm{co}}(\mathcal{A})$ such
that
\begin{equation}\label{l-s-b}
\exists\lim\limits_{n\rightarrow+\infty}F(\rho_{n})<F(\rho_{0}).
\end{equation}
As proved before for each $n=1,2,...$ there exists a measure $\mu_{n}$
in  $\mathcal{P}_{\{\rho_{n}\}}(\mathcal{A})$ such that
$F(\rho_{n})=\hat{f}(\mu_{n})$. Since the sequence $\{\rho_{n}\}$ is a compact set, the above-stated compactness criterion implies existence of a subsequence
$\{\mu_{n_{k}}\}$ converging to a particular measure $\mu_{0}$. By
continuity of the map $\mu\mapsto\mathbf{b}(\mu)$ the measure
$\mu_{0}$ belongs to the set
$\mathcal{P}_{\{\rho_{0}\}}(\mathcal{A})$. Lower semicontinuity of
functional (\ref{ess-sup-fun}) implies
$$
F(\rho_{0})\leq
\hat{f}(\mu_{0})\leq\liminf_{k\rightarrow+\infty}\hat{f}(\mu_{n_{k}})=\lim_{k\rightarrow+\infty}F(\rho_{n_{k}}),
$$
contradicting to (\ref{l-s-b}).\vspace{5pt}

The last assertion of the proposition follows from the previous ones
and Lemma 1 in \cite{H-Sh-W}. $\square$

\subsection{Existence of a state with a given finite Schmidt number such that any its countable convex decomposition does not contain pure states with a finite Schmidt rank}

In this section we show first that the separable non-countably
decomposable state constructed in \cite{H-Sh-W} has, in fact,
more stronger property: any countable decomposition of this state does not contain pure states with finite Schmidt rank
(non-countable decomposability means nonexistence of such
decomposition into product pure states -- states with the Schmidt
rank $=1$). By using this observation we construct a state with a given finite Schmidt number such that any countable decomposition of this state does not contain pure states with a finite Schmidt rank.

We present the construction of the above
mentioned separable state following the notations of \cite{H-Sh-W}. Consider the one-dimensional rotation group
represented as the interval $[0,2\pi )$ with addition
\textrm{mod}$2\pi .$ Let $\mathcal{H=} L^{2}([0,2\pi))$ with the
normalized Lebesgue measure $\frac{dx}{2\pi },$ and let $\left\{
|k\rangle ;k\in \mathbf{Z}\right\} $ be the orthonormal basis of
trigonometric functions, so that
\begin{equation*}
\langle k|\psi \rangle =\int_{0}^{2\pi }e^{-ixk}\psi
(x)\frac{dx}{2\pi }.
\end{equation*}
Consider the unitary representation $\,x\rightarrow V_{x}$ , where $
V_{x}=\sum_{-\infty }^{+\infty }e^{ixk}|k\rangle \langle k|,$ so that
$ (V_{u}\psi )(x)=\psi (x+u)$.

For arbitrary unit vectors $|\varphi _{j}\rangle \in
\mathcal{H}_{j}\simeq L^{2}([0,2\pi)),j=1,2,$ consider the separable
state
\begin{equation}
\rho _{12}=\int_{0}^{2\pi }V_{x}^{(1)}|\varphi _{1}\rangle \langle
\varphi _{1}|(V_{x}^{(1)})^*\otimes V_{x}^{(2)}|\varphi _{2}\rangle
\langle \varphi _{2}|(V_{x}^{(2)})^*\;\frac{dx}{2\pi }.
\label{cind}
\end{equation}.

The following proposition strengthens the assertion of Theorem 3 in
\cite{H-Sh-W}. It is obtained by a natural generalization of the proof of this theorem.\smallskip

\begin{property}\label{state}
 \textit{Let $\rho _{12}$ be the separable state defined in (\ref{cind}). If the vectors $|\varphi _{j}\rangle$ have  nonvanishing
Fourier coefficients then the operator
$$
\rho _{12}-\lambda\sigma
$$
is not positive for any $\lambda>0$ and any pure state $\sigma$ with
finite Schmidt rank.}\smallskip

\emph{It follows that any countable decomposition of the state $\rho _{12}$ does not contain pure states with finite Schmidt rank.}
\end{property}\smallskip

\textbf{Proof.} Suppose there exists a vector $|\psi\rangle$ in $\H_1\otimes\H_2$ with
the Schmidt rank $n$ such that
\begin{equation}
\rho _{12}\geq |\psi\rangle \langle \psi|.  \label{main-ineq}
\end{equation}
Let $|\psi\rangle=\sum_{i=1}^n
|\alpha^1_i\rangle\otimes|\alpha^2_i\rangle$, where $\{|\alpha^j_i\rangle\}_{i=1}^n$, $j=1,2$, are sets of orthogonal vectors. Inequality
(\ref{main-ineq}) implies
\begin{equation}
\int_{0}^{2\pi }\left\vert \langle \lambda _{1}|V_{x}^{(1)}|\varphi
_{1}\rangle \right\vert ^{2}\left\vert \langle \lambda
_{2}|V_{x}^{(2)}|\varphi _{2}\rangle \right\vert
^{2}\;\frac{dx}{2\pi }\geq \left|\sum_{i=1}^n
\langle\lambda_1|\alpha^1_i\rangle\langle\lambda_2|\alpha^2_i\rangle\right|^{2}
\label{b-ineq}
\end{equation}
for arbitrary $\lambda _{j}\in L^{2}([0,2\pi ))$, $j=1,2$.

Consider the linear maps
$$
L^{2}([0,2\pi))\ni\lambda
\mapsto\Phi_j(\lambda)=\{\langle\alpha^j_i|\lambda\rangle\}_{i=1}^n\in
\mathbb{C}^n
$$
and
$$
L^{2}([0,2\pi))\ni\lambda \mapsto\Psi_j(\lambda)=
\overline{\langle\lambda|V_{x}^{(j)}|\varphi _{j}\rangle} =\sum_{k=-\infty
}^{+\infty }\langle \varphi _{j}|k\rangle \langle k|\lambda \rangle
e^{-ikx},\quad j=1,2.
$$

Let $\H_0$ be a dense subset of $L^{2}([0,2\pi))$ consisting of
functions having finite number of nonzero Fourier coefficients
(trigonometric polynomials). Since $\langle\varphi
_{j}|k\rangle\neq0$ for all $k$,  the maps $\Psi_j$, $j=1,2$, are
linear isomorphisms from $\H_0$ onto itself. Hence (\ref{b-ineq})
implies
\begin{equation}
|\langle A_1(\xi), \Xi(A_2(\eta))\rangle_{\mathbb{C}^n}|^2\leq \int_{0}^{2\pi }\left\vert \xi(x)\eta(x)\right\vert ^{2}\; \frac{dx}{2\pi },\quad \xi,\eta\in\H_0, \label{b12}
\end{equation}
where
$A_j(\cdot)=\Phi_j(\Psi^{-1}_j(\cdot))$, $j=1,2$, are
linear maps from $\H_0$ to $\mathbb{C}^n$ and $\Xi$ is the complex conjugation in $\mathbb{C}^n$.

Since $\{\Phi_2(\lambda)\,|\,\lambda\in L^{2}([0,2\pi))\}=\mathbb{C}^n$, we have
$\{\Phi_2(\lambda)\,|\,\lambda\in \H_0\}=\mathbb{C}^n$ and hence
$\{A_2(\xi)\,|\,\xi\in \H_0\}=\mathbb{C}^n$. Thus there exists a subset $|\eta_1\rangle,...,|\eta_n\rangle$ of the basis $\{|k\rangle\}$
such that the vectors $A_2(\eta_1),...,A_2(\eta_n)$ form a basis in $\mathbb{C}^n$. Since $|\eta_i(x)|=1$, (\ref{b12}) implies
\begin{equation*}
|\langle A_1(\xi), \Xi(A_2(\eta_i))\rangle_{\mathbb{C}^n}|^2\leq \int_{0}^{2\pi }\left\vert \xi(x)\right\vert ^{2}\; \frac{dx}{2\pi }=\|\xi\|^2,\quad i=\overline{1,n},\quad \xi\in\H_0.
\end{equation*}
Hence the map $A_1$ is bounded on $\H_0$ and can be extended to the bounded linear operator $A_1$ from
$L^{2}([0,2\pi))$ to $\mathbb{C}^n$.

By the similar reasoning the map $A_2$ can be extended to the bounded linear operator $A_2$ from
$L^{2}([0,2\pi))$ to $\mathbb{C}^n$.

Since the anti-linear operator $B=A_1^*\circ\Xi\circ A_2$ in the space $L^{2}([0,2\pi))$
has rank $\leq n$, it can be represented as follows
$B(\cdot)=\sum_{i=1}^n\langle\cdot|\beta^2_i\rangle|\beta^1_i\rangle$, where
$\{|\beta^j_i\rangle\}$, $j=1,2$,  are sets of vectors in
$L^{2}([0,2\pi))$ and the set $\{|\beta^1_i\rangle\}$ consists of linearly
independent vectors.

Thus we can rewrite (\ref{b12}) as follows
\begin{equation}
\left|\sum_{i=1}^n\langle\xi|\beta^1_i\rangle\langle\eta|
\beta^2_i\rangle\right|^2\leq \int_{0}^{2\pi }\left\vert \xi(x)\eta(x)\right\vert ^{2}\; \frac{dx}{2\pi }. \label{b12+}
\end{equation}

By Lemma \ref{bl} below for arbitrary $\,\varepsilon>0\,$  one can find
a subset $\mathcal{A}\subset[0,2\pi)$ with the Lebesgue measure
$\,<\varepsilon\,$ such that the functions $\beta^1_1,
\beta^1_2,...,\beta^1_n$ are linearly independent on $\mathcal{A}$.
Thus for each $i$ we can find a function $\xi$ supported by
$\mathcal{A}$ such that $\langle\xi|\beta^1_i\rangle\neq0$ but
$\langle\xi|\beta^1_j\rangle=0$ for all $j\neq i$. Hence for
this function $\xi$ and arbitrary function $\eta$ supported
by the complement of $\mathcal{A}$, the right hand side of
(\ref{b12+}) vanishes implying $\langle\eta|\beta^2_i\rangle=0$
and therefore $\beta^2_i$ vanishes a.e. on the complement of
$\mathcal{A}$. It follows that the support of $\beta^2_i$ has
measure $\leq\varepsilon$, i.e. $\beta^2_i$ vanishes a.e. Thus we
have $B=0$. This shows that $|\psi\rangle=0$.
$\square $

\medskip
In the following lemma the linear independence of measurable functions $\;f_1,...,f_n\,$ on a measurable subset
$\mathcal{A}\subset\mathbb{R}$ means that any nontrivial linear combination of these functions $\neq0$ a.e. on $\mathcal{A}$.
\medskip
\begin{lemma}\label{bl}
\emph{Let $\;f_1,...,f_n\,$ be linearly independent measurable functions on
$[a,b]$. For arbitrary $\,\varepsilon>0$ there exists a subset
$\mathcal{A}\subset[a,b]$ with the Lebesgue measure
$\mu(\mathcal{A})<\varepsilon\,$ such that the functions
$\;f_1,...,f_n\,$ are linearly independent on $\,\mathcal{A}$.}
\end{lemma}
\medskip

\textbf{Proof.}  For $n=1,2$ the assertion of the lemma is obvious.
Assume it is valid for given $n$ and show its validity for $n+1$.\footnote{There is an elegant proof of this lemma non-using the induction method \cite{Shulman}.}

By the assumption for arbitrary $\varepsilon>0$ and for each family
$\{f_i\}\setminus f_j$, $j=\overline{1,n+1}$, there exists a subset
$\mathcal{A}^{\varepsilon}_j\subset[a,b]$ with
$\mu(\mathcal{A}^{\varepsilon}_j)<\varepsilon$ such that the
functions of the above family are linearly independent on
$\mathcal{A}^{\varepsilon}_j$.

If the assertion of the lemma is not valid for $n+1$ then there
exists $\varepsilon_*>0\,$ such that the functions
$\,f_1,...,f_{n+1}\,$ are linearly dependent on any subset
$\mathcal{A}\subset[a,b]\,$ with $\mu(\mathcal{A})<\varepsilon_*$.

Let $\varepsilon<\varepsilon_*/2(n+1)$ and
$\mathcal{A}^{\varepsilon}=\bigcup_{j=1}^{n+1}\mathcal{A}^{\varepsilon}_j$.
Choose a finite collection  $\{\mathcal{B}_k\}$ of disjoint subsets of
$[a,b]\setminus\mathcal{A}^{\varepsilon}$ such that
$\mu(\mathcal{B}_k)<\varepsilon_*/2$ and
$\bigcup_k\mathcal{B}_k=[a,b]\setminus\mathcal{A}^{\varepsilon}$.

For each $k$ let
$\mathcal{C}_k=\mathcal{A}^{\varepsilon}\cup\mathcal{B}_k$. Since
$\mu(\mathcal{C}_k)<\varepsilon_*$ there exists a set
$\{\lambda^k_i\}_{i=1}^{n+1}$ of complex numbers such that
\begin{equation}\label{l-c}
\sum_{i=1}^{n+1}\lambda^k_i f_i(x)=0\quad \text{a.e. on}
\;\,\mathcal{C}_k, \quad
\sum_{i=1}^{n+1}|\lambda_i^{k}|>0.
\end{equation}
Since $\mathcal{A}^{\varepsilon}_j\subset\mathcal{C}_k$ for all $j$,
it is easy to see that $\lambda^k_i\neq0$ for all $i$. So, we may
assume that $\lambda^k_{n+1}=1$. Since the functions
$\,f_1,...,f_{n+1}\,$ are linearly independent on
$[a,b]=\bigcup_k\mathcal{C}_k$, there exist $k_1$ and $k_2$ such
that
$\{\lambda^{k_1}_i\}_{i=1}^{n+1}\neq\{\lambda^{k_2}_i\}_{i=1}^{n+1}$.
It follows from (\ref{l-c}) that
$$
\sum_{i=1}^{n}(\lambda_i^{k_1}-\lambda_i^{k_2}) f_i(x)=0\quad \text{a.e. on}\;\,\mathcal{A}_{n+1}^{\varepsilon}\subseteq\mathcal{C}_{k_1}\cap\mathcal{C}_{k_2},\qquad
\sum_{i=1}^{n}|\lambda_i^{k_1}-\lambda_i^{k_2}|>0
$$
contradicting to the construction of the set
$\mathcal{A}_{n+1}^{\varepsilon}$. $\square$ \medskip

\noindent\textbf{Example of a state $\,\omega$ with $\,\sn(\omega)=k\in\mathbb{N}$ such that the
operator $\,\omega-\lambda\sigma$ is not positive for any pure state $\sigma$ with
a finite Schmidt rank and any $\lambda>0$.}\smallskip

Let $\{|\varphi^{i}_{1}\rangle\}_{i=1}^k$ and
$\{|\varphi^{i}_{2}\rangle\}_{i=1}^k$ are collections of orthogonal
unit vectors in $\H_1=L^{2}([0,2\pi ))$ and $\H_2=L^{2}([0,2\pi ))$
correspondingly with non-vanishing Fourier coefficients. Let $\K$ be
the $k$-dimensional Hilbert space with the orthonormal basis $\{|i\rangle\}_{i=1}^k$. For each natural $n$ consider the
state
\begin{equation}
\rho^n_{123}=\int_{0}^{2\pi/n}V_{x}^{(1)}\otimes V_{x}^{(2)}\otimes
I_{\K}\,\cdot|\Omega\rangle \langle \Omega|\cdot\,(V_{x}^{(1)})^*\otimes
(V_{x}^{(2)})^*\otimes I_{\K}\;\frac{ndx}{2\pi } \label{cind+}
\end{equation}
in $\S(\H_1\otimes\H_2\otimes\K)$, where
$|\Omega\rangle=\frac{1}{\sqrt{k}}\sum_{i=1}^k
|\varphi^{i}_{1}\rangle\otimes|\varphi^{i}_{2}\rangle\otimes|i\rangle$
is a unit vector in $\H_1\otimes\H_2\otimes\K$.\smallskip

In what follows (speaking about the Schmidt rank and the Schmidt number) we will treat the space $\H_1\otimes\H_2\otimes\K$ as the tensor product of the spaces
$\H_1$ and $\H_2\otimes\K$.\smallskip

Since the state $\rho^n_{123}$ belongs to the convex closure of the
family of local unitary translations of the state
$|\Omega\rangle\langle\Omega|$ such that
$\sr(|\Omega\rangle\langle\Omega|)=k$, we have $\sn(\rho^n_{123})\leq k$
for all $n$. Since the sequence $\{\rho^n_{123}\}$ tends to the
state $|\Omega\rangle\langle\Omega|$, this and lower semicontinuity of the
Schmidt number (Proposition \ref{sn-def-p}A) show that $\sn(\rho^n_{123})=k$ for  sufficiently
large $n$.

Suppose that
\begin{equation}
\rho^n_{123}\geq \lambda|\Psi\rangle\langle\Psi|, \label{main-ineq+}
\end{equation}
for some $\lambda>0$, where $|\Psi\rangle\langle\Psi|$ is a pure
state in $\S(\H_1\otimes\H_2\otimes\K)$ with a finite Schmidt rank.
Let $P_i=I_{\H_1}\otimes I_{\H_2}\otimes|i\rangle\langle i|$. Since
$\sum_{i=1}^k P_i=I_{\H_1\otimes\H_2\otimes\K}$, there exists $i_0$
such that $P_{i_0}|\Psi\rangle\neq0$. We may assume that $i_0=1$. Hence
$P_1|\Psi\rangle=\nu|\psi\rangle\otimes|1\rangle$, where $\nu>0$ and $|\psi\rangle$
is a unit vector in $\H_1\otimes\H_2$.

Since $P_1\rho^n_{123}P_1=k^{-1}\rho^n_{12}\otimes|1\rangle\langle1|$,
where
\begin{equation*}
\rho^n_{12}=\int_{0}^{2\pi/n}V_{x}^{(1)}|\varphi^{1}_{1}\rangle
\langle \varphi^{1}_{1}|(V_{x}^{(1)})^*\otimes
V_{x}^{(2)}|\varphi^{1}_{2}\rangle \langle
\varphi^{1}_{2}|(V_{x}^{(2)})^*\;\frac{ndx}{2\pi},
\end{equation*}
it follows from (\ref{main-ineq+}) that
$\rho^n_{12}\geq k\lambda\nu|\psi\rangle\langle\psi|$. Since
$P_1(\cdot)P_1$ and $\Tr_{\K}(\cdot)$ are local operations, the state
$|\psi\rangle\langle\psi|$ has finite  Schmidt rank. Hence
Proposition \ref{state} implies $\lambda=0$. $\square$
%\end{example}\smallskip

\vspace{15pt}

I am grateful to A.S.Holevo for the valuable help and to T.V.Shulman for the useful discussion.


\begin{thebibliography}{99}%{Литература}

\bibitem{ES} C.D.Aliprantis, K.C.Border, "Infinite dimensional analysis", Springer Verlag, 2006.

\bibitem{Bill} P.Billingsley, "Convergence of probability
measures",  Wiley, New York-London-Sydney, 1968.

\bibitem{p-e-b-ch} D.Chruscinski, A.Kossakowski, "On Partially Entanglement Breaking Channels", Open Sys. Information Dyn., 13, P.17-26, 2006; arXiv:quant-ph/0511244.

\bibitem{H-SSQT} A.S.Holevo, "Statistical structure of quantum theory",
Springer-Verlag, 2001.

\bibitem{H-Sh-W} A.S.Holevo, M.E.Shirokov, R.F.Werner,
"On the notion of entanglement in Hilbert spaces",  Russian Math. Surveys, 60, N.2, P.359-360, 2005; arXiv:quant-ph/0504204.

\bibitem{Sh-H} A.S.Holevo, M.E.Shirokov, "On approximation of quantum channels",  Problems of Information Transmission,
44, P.3-22, 2008; arXiv:0711.2245.

\bibitem{e-b-ch} M.Horodecki, P.W.Shor, M.B.Ruskai, "General Entanglement Breaking Channels", Rev. Math. Phys., 15, P.629-641, 2003; arXiv:quant-ph/0302031.

\bibitem{N&Ch} M.A.Nielsen, I.L.Chuang,  "Quantum Computation and Quantum
Information", Cambridge University Press, 2000.

\bibitem{J&T} A.D.Joffe,  W.M.Tikhomirov,  "Theory of extremum
problems", AP, NY, 1979.

\bibitem{Par} K.Parthasarathy, "Probability measures on metric
spaces", Academic Press, New York and London, 1967.

\bibitem{R}  R.Rockafellar,  "Convex analysis",  Tyrrell, 1970.

\bibitem{SBL} A.Sanpera, D.Brub, M.Lewenstein, "Schmidt number witnesses and bound entanglement", Phys. Rev. A 63, 050301, 2001;
arXiv:quant-ph/0009109, 2000.

\bibitem{Sh-9} M.E.Shirokov, "The properties of the set of quantum states and their application to construction of entanglement monotones", Izvestiya: Mathematics, 74:4, P.849-882, 2010; arXiv:0804.1515.

\bibitem{Sh-14} M.E.Shirokov,  "Monotonicity of the Holevo quantity: a necessary condition for equality in terms of a channel and its applications",
arXiv:1106.3297.

\bibitem{Shulman} T.V.Shulman,  private communication.

\bibitem{S&V-1} J.Sperling, W.Vogel, "The Schmidt number as an universal entanglement measure", Phys. Scr., 83, 045002, 2011;
arXiv:0908.3974, 2009.

\bibitem{S&V-2} J.Sperling, W.Vogel, "Determination of the Schmidt number",   Phys. Rev. A 83, 042315, 2011; arXiv:1103.1287.

\bibitem{T&H} B.M.Terhal, P.Horodecki, "A Schmidt number for density matrices", Phys. Rev. A Rapid Communications, 61, 040301, 2000;
arXiv:quant-ph/9911117, 2000.

\end{thebibliography}
\end{document}